\documentclass[aps,amsmath,amssymb,nofootinbib,preprint]{revtex4-1}

\usepackage{graphicx}
\usepackage{dcolumn}
\usepackage{bm}
 \usepackage{url}
 \usepackage[utf8]{inputenc}
\usepackage[T1]{fontenc}
\usepackage{mathptmx}
\usepackage{gensymb}
\usepackage{etoolbox}
 \usepackage{color}
 \usepackage{lipsum}
 \usepackage[normalem]{ulem}

\begin{document}


\newpage
\clearpage
\title{High-density Optical Quantum Sensors with Pulsed Probe Read-out for Correlated Spin-Noise Reduction }
\author{Igor Savukov$^{\mathsection}$}
\email{isavukov@lanl.gov.}
\author{Young Jin Kim$^{\mathsection}$}
\email{youngjin@lanl.gov.}

\noindent LA-UR-26-24800

\affiliation{MPA-Quantum, Los Alamos National Laboratory, P.O. Box 1663, Los Alamos, New Mexico 87545, USA}
\def\thefootnote{$\mathsection$}\footnotetext{these authors are co-first authors who contributed equally to this work.}
\date{\today}

\begin{abstract}

Optical quantum sensors based on alkali-metal atoms enable highly sensitive magnetic-field measurements at room temperature. Their ultimate performance is limited by intrinsic spin noise once technical noise sources are sufficiently suppressed. Reducing and characterizing this noise is therefore essential for improving sensor sensitivity and for investigating quantum-enhanced sensing protocols. Here we investigate correlated spin fluctuations in a radio-frequency optical quantum sensor based on a high-density potassium vapor cell using a pulsed probe readout scheme. The system employs orthogonal pump and probe laser beams, while a static magnetic field defines the sensing frequency and synchronizes the probe pulses with the spin precession. To detect weak spin correlations, we implement a measurement protocol that combines periodic probing, phase cycling, and window-shifted acquisition. Phase cycling suppresses reproducible probe-induced coherent transients, while subtraction of temporally correlated signals acquired within the spin-relaxation time reduces the measured spin-noise level. The observed noise reduction is consistent with temporal spin correlations, including those expected in spin-squeezing protocols, although the present measurements do not uniquely distinguish this interpretation from other correlated-noise mechanisms. The demonstrated measurement protocol provides a practical route toward operating warm-vapor optical quantum sensors closer to their fundamental spin-noise limit and may benefit radio-frequency magnetometry and sensing of weak coherent signals, including emerging dark-matter detection schemes.

\end{abstract}

\maketitle

\section*{Introduction}
Fundamental quantum spin noise, arising from intrinsic fluctuations of spin ensembles, has been studied in a variety of physical systems, including semiconductor spin-noise spectroscopy~\cite{Crooker2004,Oestreich2005} and atomic ensembles probed using quantum non-demolition (QND) measurements, which provide a proven pathway for generating spin-squeezed states~\cite{Hammerer2010,Kuzmich1997,Kuzmich_1998}. In many precision measurements, spin noise sets a fundamental limit to sensitivity. Reducing or controlling these intrinsic fluctuations is therefore a central challenge in quantum sensing. One promising approach is spin squeezing, a QND-based technique that redistributes quantum uncertainty among spin components and is closely related to entanglement in collective spin systems~\cite{Kitagawa1993,Wineland1994}. Spin squeezing has been demonstrated in cold atomic ensembles and optical cavities~\cite{Kuzmich1997,Julsgaard2001,Appel2009,Wasilewski2010,Esteve2008,Gross2010,Riedel2010}, where it enables measurements beyond the standard quantum limit (SQL)~\cite{Wineland1994,Pezze2018}. More recently, quantum-enhanced magnetometry beyond the projection-noise limit has been demonstrated using spin-squeezed atomic ensembles~\cite{Koschorreck2010,Sewell2012}, while large metrological gains have also been achieved in systems relevant for precision sensing and atomic clocks~\cite{Hosten2016}. Extending these concepts to practical sensing platforms, particularly warm-vapor alkali-vapor sensors, remains an important challenge.

Overcoming the SQL is particularly important for ultra-sensitive magnetic-field measurements, where sensor performance is often limited by fundamental noise sources. Developing measurement protocols capable of reducing or exploiting correlated spin fluctuations in realistic sensing systems would therefore represent an important step toward improving the sensitivity of atomic magnetometers. High-sensitivity optical quantum sensors (OQSs) of atomic magnetometry, capable of reaching sub-fT sensitivity, are important for applications involving extremely weak magnetic fields, ranging from fundamental physics~\cite{PhysRevLett.121.091802,NatureCom_Kim,PhysRevD.108.052007} to biomagnetic imaging~\cite{SAVUKOV2009188,APL2019}. These sensors operate at room temperature using vapor cells filled with alkali-metal atoms and laser light: a circularly polarized pump beam prepares a spin-polarized state, while a linearly polarized probe beam detects spin precession induced by external static magnetic fields via optical rotation~\cite{Budker2007}. Recent advances in warm-vapor atomic magnetometry have approached fundamental sensitivity limits, motivating renewed interest in measurement protocols capable of suppressing intrinsic spin noise and exploiting correlated spin dynamics in practical sensing systems~\cite{Budker2007,Shah2007}.

In such systems, once technical laser and environmental noise sources are sufficiently suppressed, intrinsic spin noise becomes a dominant fundamental limitation~\cite{Budker2007}. To illustrate this, we consider a radio-frequency (RF) OQS~\cite{SavukovRFMag}, for which the magnetic-field sensitivity can be expressed as
\begin{equation}
dB=\frac{1}{\gamma \sqrt{nV}}\sqrt{\frac{4}{T_2}+\frac{R_{\mathrm{pr}}\text{OD}}{32}+\frac{8}{R_{\mathrm{pr}}\text{OD}T_2^2\eta}},
\end{equation}
where the terms represent contributions from intrinsic spin noise, probe-induced back-action, and photon shot noise. Here $\gamma$ is the gyromagnetic ratio of alkali-metal atoms, $n$ is the spin density inside a vapor cell, $V$ is the active measurement cell volume, $T_2$ is the transverse spin relaxation time, $R_{\mathrm{pr}}$ is the probe-induced spin-destruction rate, $\text{OD}$ is the optical depth of the probe beam inside the cell, and $\eta$ is the photodetector quantum efficiency in the spin state readout. Optimizing the product $R_{\mathrm{pr}}\mathrm{OD}$,
\begin{equation}
R_{\mathrm{pr}}\mathrm{OD}=16/(T_2\sqrt{\eta})
\label{eq2}
\end{equation}
yields a minimum sensitivity
\begin{equation}
dB^{\min}=\frac{1}{\gamma \sqrt{nV}}\sqrt{\frac{4}{T_2}\left(1+\frac{1}{4\sqrt{\eta}}\right)},
\end{equation}
showing that, in the optimized regime, the sensitivity is primarily limited by the spin noise. The relative importance of spin noise increases with the spin relaxation time $T_2$, making it particularly relevant in systems with long coherence times. 

In OQSs, the probe beam intensity plays a dual role. Increasing probe beam power reduces photon shot noise but also introduces additional spin relaxation through probe-induced spin destruction. As a result, the transverse spin relaxation rate is not independent of the probe, but increases with probe intensity. In addition, residual magnetic-field gradients contribute an extra dephasing rate that can be comparable to the intrinsic spin-relaxation rate. These effects modify the optimal operating point of the sensor. Therefore, the optimal probe intensity is determined by a trade-off between photon shot noise and total spin relaxation, including both intrinsic and probe-induced contributions, as well as gradient broadening. In this work, we account for these effects when estimating the optimal probe conditions.

Spin squeezing provides one possible route toward reducing spin noise by redistributing quantum uncertainty among spin components and generating entanglement within the spin ensemble~\cite{Kitagawa1993,Wineland1994,Pezze2018}. More generally, temporal correlations of collective spin fluctuations can provide additional information about the underlying spin dynamics and may enable measurement protocols that operate closer to the fundamental spin-noise limit. Investigation of such correlated spin fluctuations is therefore of considerable interest for warm-vapor OQSs. In RF OQSs, however, an additional challenge arises from the presence of a bias magnetic field used to tune the sensor to the measurement frequency. This field causes the collective spin to precess, leading to a time-dependent measurement axis. As a result, the squeezed and anti-squeezed components periodically rotate into the measurement direction, and the net squeezing effect can average out unless the measurement is synchronized with the spin precession. To investigate correlated spin fluctuations in RF OQSs, we develop and experimentally implement a pulsed probe measurement protocol synchronized with the spin precession. The approach combines pulsed readout, phase cycling, and time-correlated acquisition to exploit temporal spin correlations over the characteristic relaxation time $T_2$. These techniques suppress coherent probe-induced artifacts while reducing the measured spin-noise level. The resulting protocol provides a practical approach for operating RF OQSs closer to their intrinsic spin-noise limit and establishes an experimental framework for future investigations of quantum-enhanced sensing, including spin squeezing.

These methods are compatible with pulsed detection schemes used in nuclear magnetic resonance (NMR) and nuclear quadrupole resonance (NQR)~\cite{KIM201435}, where phase cycling is routinely employed. They are also relevant for emerging low-mass dark matter detection schemes in which weak RF magnetic signals are coupled to high-$Q$ resonant circuits and detected with sensitive magnetometers~\cite{PhysRevD.108.052007}. In such systems, fluctuating collective spin polarization generates magnetic-field noise that couples to the detection resonant circuit, which is amplified by the resonator by a factor of $Q$. For sufficiently high $Q$, this contribution can become significant and limit sensitivity. Suppression of both probe-induced correlations and intrinsic spin noise is therefore important for achieving optimal performance. The present protocol provides a pathway to reducing these noise contributions while maintaining sensitivity to externally generated signals with finite coherence time.

Although pulsed probing and phase cycling are individually well-established
techniques in optical magnetometry, quantum non-demolition (QND)
measurements, and NMR~\cite{Budker2007,NatPhysBudker2007,Shah2007,ShahPRL2010,Koschorreck2010,KIM201435},
their combined use for investigating delay-dependent correlated spin
fluctuations in a high-density RF optical quantum sensor has not, to our
knowledge, been previously demonstrated. Previous work has primarily focused
on continuous QND measurements, spin squeezing, or high-bandwidth optical
magnetometry~\cite{ShahPRL2010,Koschorreck2010,Sewell2012}, whereas the
present work exploits synchronized pulsed probing, phase cycling, and
variable-delay window subtraction to investigate temporal spin correlations
while suppressing coherent probe-induced artifacts. The present work does not
introduce a new phase-cycling method or a new pulsed-probe technique
individually; rather, it combines these established techniques into a
measurement protocol applicable to practical warm-vapor RF optical
magnetometers. This protocol provides a practical framework for operating RF
OQSs closer to their intrinsic spin-noise limit and for future investigations
of quantum-enhanced sensing based on correlated spin fluctuations.

\section*{Results}
\subsection*{Measurement protocol and experimental concept.} 
To investigate correlated spin fluctuations using the pulsed probe protocol, we designed a measurement scheme synchronized with the spin dynamics of RF OQSs. Because spin fluctuations remain correlated over timescales shorter than $T_2$, sequential probe pulses separated by controlled delays can sample partially correlated spin noise. By appropriately shifting the acquisition window and subtracting consecutive measurements within this correlation time, noise contributions can be reduced while preserving the underlying spin-noise signal. Unlike conventional QND squeezing experiments performed in a static quantization axis, the present protocol is adapted to an RF OQS, where Larmor precession continuously rotates the measurement basis. In addition, phase cycling is employed to suppress reproducible probe-induced transients between experimental runs. Since fundamental spin fluctuations are uncorrelated between runs, this subtraction removes technical artifacts without eliminating the intra-run correlations used to detect spin-noise reduction.

The performance of this approach depends on operating in a regime where intrinsic spin noise is significant while probe-induced perturbations remain controlled. To estimate relevant parameters, we consider a probe beam with a diameter of $0.5$ mm and a path length of $1$ cm inside a potassium vapor cell. At a cell temperature of $169^\circ\mathrm{C}$, the atomic density is approximately $3.53 \times 10^{13}$~$\mathrm{cm}^{-3}$, corresponding to an optical depth of $\mathrm{OD} \approx 31$, based on scaling from Ref.~\cite{SavukovRFMag}.
Under the optimized probe conditions in Eq.~(\ref{eq2}), the probe-induced spin-destruction rate is estimated as
\begin{equation}
R_{\mathrm{pr}} \approx \frac{16}{T_2\mathrm{OD}\sqrt{\eta}},
\end{equation}
yielding $R_{\mathrm{pr}} \approx 0.73/T_2$ $\mathrm{s}^{-1}$ for typical parameters. This rate is smaller than the collisional spin-relaxation rate, indicating that probe-induced decoherence remains limited. Including spin-exchange effects modifies the effective relaxation rate and leads to an increased optimal probe rate on the order of $R_{\mathrm{pr}} \sim 100$ $\mathrm{s}^{-1}$.

\subsubsection*{Probe optimization including relaxation and gradient broadening}

To estimate the optimal probe conditions, we include the dependence of the transverse relaxation rate on probe-induced spin destruction as well as additional dephasing due to magnetic-field gradients. In the spin-exchange-narrowed regime, the intrinsic spin relaxation rate can be approximated as
\begin{equation}
\frac{1}{T_2} \approx \sqrt{\frac{R_{\mathrm{SE}}\left(R_{\mathrm{SD}} + R_{\mathrm{pr}}\right)}{5}},
\end{equation}
where $R_{\mathrm{SE}}$ is the spin-exchange rate and $R_{\mathrm{SD}}$ is the spin-destruction rate.
To account for magnetic-field gradients, we introduce an additional dephasing rate $\Gamma_{\mathrm{grad}}$ and define the effective transverse relaxation rate as
\begin{equation}
\Gamma_{\mathrm{eff}}(R_{\mathrm{pr}}) =
\sqrt{\frac{R_{\mathrm{SE}}\left(R_{\mathrm{SD}} + R_{\mathrm{pr}}\right)}{5}}
+ \Gamma_{\mathrm{grad}}.
\end{equation}
Then, the magnetic-field sensitivity can then be written as
\begin{equation}
dB = \frac{1}{\gamma \sqrt{nV}}
\sqrt{
4 \Gamma_{\mathrm{eff}} +
\frac{R_{\mathrm{pr}} \, \mathrm{OD}}{32} +
\frac{8 \Gamma_{\mathrm{eff}}^2}{R_{\mathrm{pr}} \, \mathrm{OD} \, \eta}
}.
\end{equation}
In this formulation, the probe rate $R_{\mathrm{pr}}$ affects the sensitivity in two competing ways: increasing $R_{\mathrm{pr}}$ reduces photon shot noise, while simultaneously increasing the effective relaxation rate $\Gamma_{\mathrm{eff}}$. Therefore, the optimal probe rate must be obtained by minimizing $dB$ with respect to $R_{\mathrm{pr}}$. In contrast to the case where $T_2$ is treated as independent of probe power, this optimization cannot be performed analytically and is evaluated numerically. When the gradient-induced broadening is comparable to the intrinsic relaxation rate, $\Gamma_{\mathrm{grad}} \sim \sqrt{R_{\mathrm{SE}}R_{\mathrm{SD}}/5}$, the optimal probe rate is primarily determined by the total effective linewidth rather than the intrinsic $T_2$ alone. These conditions place the system in a regime where spin noise is a dominant contribution while maintaining manageable probe back-action, making it suitable for investigating correlated spin fluctuations and possible spin-noise reduction.
For representative parameters corresponding to potassium vapor at $169^\circ\mathrm{C}$, 
with $n \approx 3.5 \times 10^{13}\,\mathrm{cm^{-3}}$, $\mathrm{OD} \approx 30$, 
and $\eta \approx 0.7$, we obtain an optimal probe-induced spin-destruction rate 
$R_{\mathrm{pr}} \sim 30\text{--}50~\mathrm{s^{-1}}$ when gradient broadening is included. 
This is significantly reduced compared to the estimate obtained for fixed $T_2$.

\subsection*{Experimental setup.} 

\begin{figure}
\includegraphics[width=6 in]{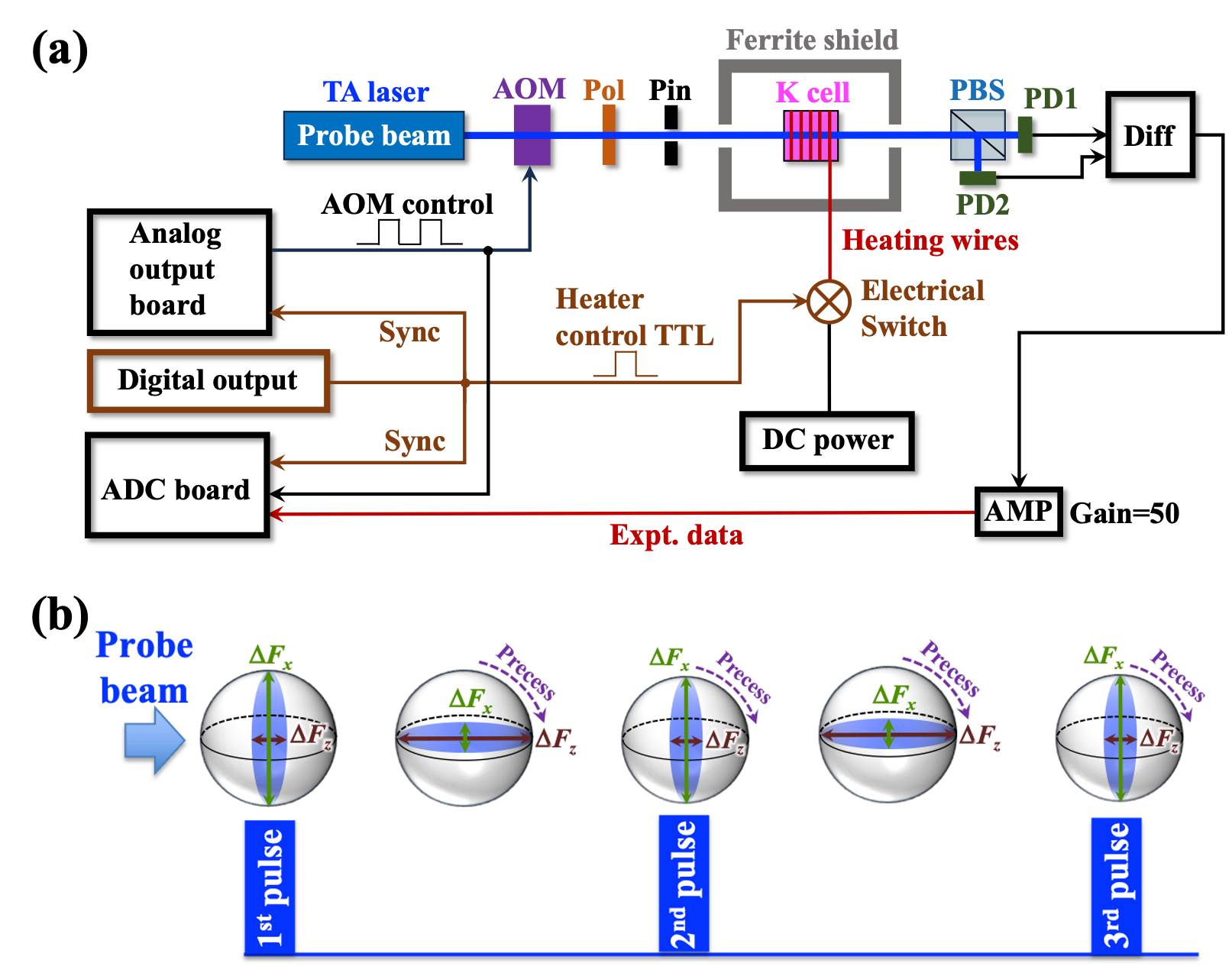}
\caption{\textbf{Setup for correlated spin noise measurement.} (a) Schematic of the experimental setup using a pulsed probe beam, where 
TA: tapered amplifier (probe laser); 
AOM: acousto-optic modulator; 
Pin: pin hole;
Pol: polarizer; 
K cell: potassium vapor cell; 
PBS: polarizing beam splitter; 
PD: balanced photodetector;
Diff: Transimpedance amplifier and PD difference;
AMP: voltage amplifier with the gain of 50;
(b) Pulsed probe measurements at selected phases of the spin precession. The first pulse provides a reference measurement, while subsequent pulses sample the same spin component after controlled delays to probe temporal spin correlations.}
\label{fig:setup}
\end{figure}

We developed an RF OQS test module based on a potassium vapor cell and orthogonal pump--probe laser beams. To suppress ambient magnetic-field noise, the vapor cell is enclosed within a cylindrical ferrite shield. A schematic of the experimental setup is shown in Fig.~\ref{fig:setup}(a).
To implement pulsed probe readout, an acousto-optic modulator (AOM) is placed after the probe laser tapered amplifier, enabling modulation of the probe beam at frequencies up to 1~MHz. As illustrated in Fig.~\ref{fig:setup}(b), the AOM is used to generate probe pulses synchronized with the spin precession, such that measurements are performed at selected phases of the spin precession where temporal spin correlations are expected to be strongest.
This requires the probe pulse sequence to be synchronized with twice the Larmor precession frequency. The factor of two arises because the measured spin projection is unchanged
after a $180^\circ$ rotation of the transverse spin polarization. Successive
probe pulses therefore sample the same spin component every half Larmor
period, maximizing the temporal correlation between consecutive measurements.
Synchronization at twice the Larmor frequency ensures that each probe pulse
interrogates the same spin projection despite the continuous spin precession. The probe beam passes through the AOM and a polarizer before traversing the potassium vapor cell. After the cell, the transmitted beam is analyzed using a polarizing beam splitter (PBS), and the two output ports are detected by balanced photodetectors. The resulting photocurrents are converted to voltages using a custom transimpedance amplifier and subtracted electronically to produce a differential signal. Precise balancing is achieved by fine adjustment of the input polarizer angle, which is essential for suppressing noise arising from probe intensity fluctuations and AOM-induced modulation. The probe beam thus provides a sensitive readout of the collective spin state of the potassium atoms in the vapor cell.

The cell temperature is maintained at approximately $180^\circ\mathrm{C}$ by resistive heating, in which non-magnetic tungsten wire, connected to a DC power supply via an electrical switch, is wrapped around the outer surface of the vapor cell.
To eliminate magnetic fields and noise generated by the DC heating current during measurements, the heating current is periodically switched off. A stabilized power supply provides the heating current, and the switch controls the on--off cycle. The measurements are performed using a data acquisition system (DAQ, NI PXIe-4480), which includes an analog output board and a 24-bit analog-to-digital converter (ADC) board, and are synchronized with the heating cycle using TTL pulses generated by a digital output of the analog output board. These TTL pulses trigger both the analog output board, which generates the experimental probe pulse sequences, and the ADC board.
The DAQ sampling rate is set sufficiently high to minimize artifacts associated with probe pulsing and to ensure stable and accurate noise characterization.

A circularly polarized pump beam is applied orthogonally to the probe beam, as in a standard OQS configuration. For simplicity, this subsystem is not shown in Fig.~\ref{fig:setup}(a).
The pump beam is not modulated and is tuned to maximize the OQS signal. Data acquisition and processing are performed using the DAQ with a home-built LabVIEW program.
The probe pulse sequences defined in the program are generated by the analog output board and simultaneously recorded by the ADC board to ensure accurate synchronization and to eliminate hidden timing delays.

\begin{figure}
\includegraphics[width=6.5 in]{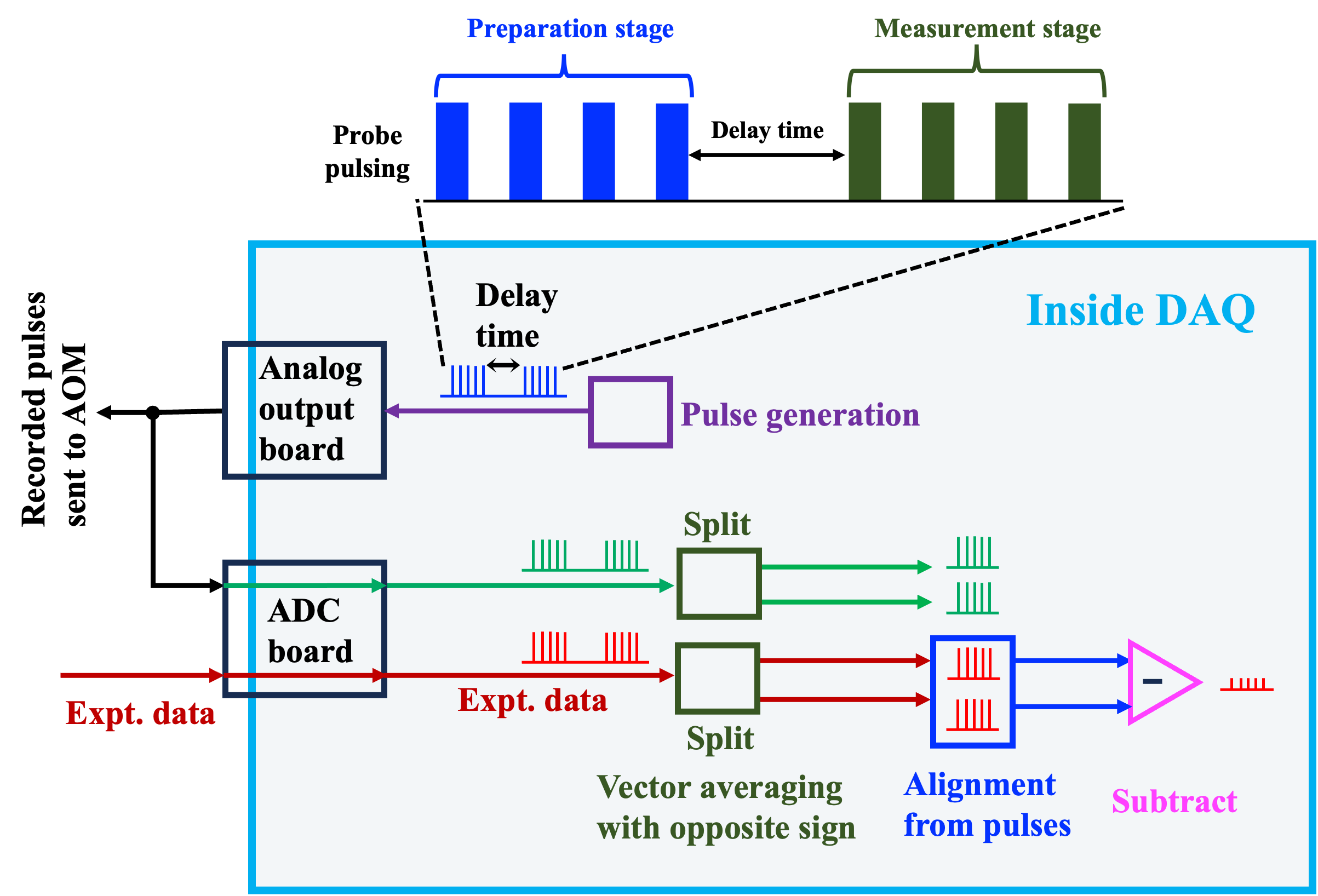}
\caption{\textbf{Block diagram of the LabVIEW processing program.} Two probe pulse trains are implemented to prepare the spin-correlated state and perform the spin noise measurements, which are split by the delay time. They are temporally aligned and synchronously subtracted to implement the correlated spin-noise measurement protocol. }
\label{fig:labview fig}
\end{figure}

Two probe pulse trains are implemented, as illustrated in
Fig.~\ref{fig:labview fig}. The first pulse train serves as a reference
measurement, while the second measures the same spin component after a
controlled delay. By varying this delay, the temporal evolution of the
correlated spin-noise signal is measured. The resulting delay dependence is
used to extract the characteristic transverse spin-relaxation time $T_2$ and
to investigate temporal spin correlations.
The two pulse trains are split, temporally aligned, and synchronously subtracted to implement the pulsed correlated spin-noise measurement protocol. A key element of this approach is the subtraction of consecutive waveforms, which suppresses correlated noise and probe-induced spin excitations arising from an effective light-shift field, acting similarly to a magnetic field along the probe beam direction.
Although the probe beam is nominally linearly polarized, residual circular polarization can arise due to birefringence in the vapor cell windows. In addition, propagation through the spin-polarized vapor can enhance circular polarization components via optical pumping selection rules. We experimentally observe such probe-induced signals and efficiently suppress them by subtracting the successive waveforms. This subtraction does not reduce uncorrelated noise contributions, such as probe-induced spin-destruction noise.

\subsection*{Operating regime and limitations.}

Observation of fundamental spin noise and correlated spin dynamics requires careful optimization of probe parameters. As an initial approach, we investigated probe pulsing with a 50$\%$ duty cycle. The probe beam diameter was reduced to approximately $0.5$ $\mathrm{mm}$ to enhance the spin noise signal, and the probe was detuned from the D1 transition to reduce probe-induced spin destruction.
Under these conditions, the estimated spin noise was on the order of a few $\mathrm{fT}/\mathrm{Hz}^{1/2}$, comparable to photon shot noise. However, despite operating in a regime where $R_{\mathrm{pr}} < R_{\mathrm{SD}}$, significant probe-induced perturbations were observed. In particular, the signal exhibited a strong dependence on the delay between probe pulse trains, consistent with probe-induced spin excitations and light-shift effects.
At a 50$\%$ duty cycle, these perturbations exceeded the expected spin noise signal and obscured potential squeezing signatures. Reducing the duty cycle to approximately 20$\%$ decreased the disturbance by about an order of magnitude, enabling operation in a regime where spin noise could be more reliably detected.
The other parameters of the K cell and lasers are listed in Table I.
\begin{table*}[ht]
\caption{Principal experimental parameters used for the correlated spin-noise
measurements presented in Fig.~\ref{fig:40 avr data}. Quantities marked
"estimated" were inferred from independent measurements or operating
conditions rather than measured directly during the spin-noise experiment.}
\label{tab:exp_parameters}
\begin{ruledtabular}
\begin{tabular}{ll}
Parameter & Value \\
\hline
Alkali vapor & Potassium (K) \\
Cell dimensions & $2\times2\times2~\mathrm{cm^3}$ \\
Cell temperature (Fig.~3) & $168^\circ$C \\
Maximum operating temperature & $\sim180^\circ$C \\
Buffer gas 1 & He, $\sim1$ atm \\
Buffer gas 2 & N$_2\sim 30$ Torr \\
Pump wavelength & Near K D$_1$, adjusted for maximum OQS signal \\
Pump polarization & Circular \\
Pump power & $\sim20$ mW \\
Pump beam diameter & $\sim1$ cm \\
Probe wavelength (vacuum) & $769.60$ nm \\
Probe detuning & $+257$ GHz from the K D$_1$ resonance \\
Probe polarization & Linear \\
Probe beam diameter & $\sim0.5$ mm \\
Probe power before cell & $\sim100$ mW (estimated) \\
Photodiode signal & $300$--$600$ mV per detector ($1~\mathrm{k}\Omega$ transimpedance) \\
RF resonance frequency & $\sim35$ kHz \\
Bias magnetic field & Tuned to the RF resonance ($\gamma\approx700$ kHz/G) \\
Probe duty cycle & $20\%$ \\
Acquisition window & $\sim1$ ms \\
Measured transverse relaxation time & $T_2=0.62\pm0.10$ ms \\
Probe-induced relaxation rate & $R_{\rm pr}\approx193~\mathrm{s^{-1}}$ \\
\end{tabular}
\end{ruledtabular}
\end{table*}

\subsection*{Optimized operating conditions.}

After reducing the duty cycle to 20$\%$, the probe laser was retuned closer to the D$_1$ resonance, with a measured vacuum wavelength of approximately
769.60~nm, corresponding to a blue detuning of approximately 257~GHz from the unperturbed D$_1$ line center. During each probe pulse, the probe-induced spin-destruction rate was estimated to be
$R_{\mathrm{pr}}^{(\mathrm{pulse})}\approx193~\mathrm{s}^{-1}$.
However, because the probe duty cycle was 20\%, the corresponding time-averaged probe-induced relaxation rate was approximately
\[
\overline{R}_{\mathrm{pr}}
\approx0.2\,R_{\mathrm{pr}}^{(\mathrm{pulse})}
\approx39~\mathrm{s}^{-1},
\]
which is close to the theoretically estimated optimum range of
30--50~$\mathrm{s}^{-1}$.
Including spin-exchange and spin-destruction processes, the effective transverse relaxation rate was
\begin{equation}
T_2^{-1}\approx10^3~\mathrm{s}^{-1},
\end{equation}
corresponding to a coherence time of approximately
$1~\mathrm{ms}$. Thus, the instantaneous probe intensity provided a sufficiently strong optical rotation signal, while the duty-cycle-averaged probe-induced relaxation remained close to the optimum operating regime.
Atomic diffusion modifies the effective interaction region. The diffusion length $\sqrt{D T_2}$ with $D$ being the diffusion coefficient of pottasium atoms in the vapor cell, is approximately $0.3~\mathrm{mm}$, increasing the effective probe diameter to about $0.9~\mathrm{mm}$. This reduces the impact of local probe-induced spin destruction by distributing it over a larger volume.
Under these conditions, the ratio of spin noise to probe photon shot noise is estimated to be close to unity, placing the system near the optimal sensitivity regime.

\subsection*{Observation of correlated spin-noise reduction.}

\begin{figure}
\centering
\includegraphics[width=5.5 in]{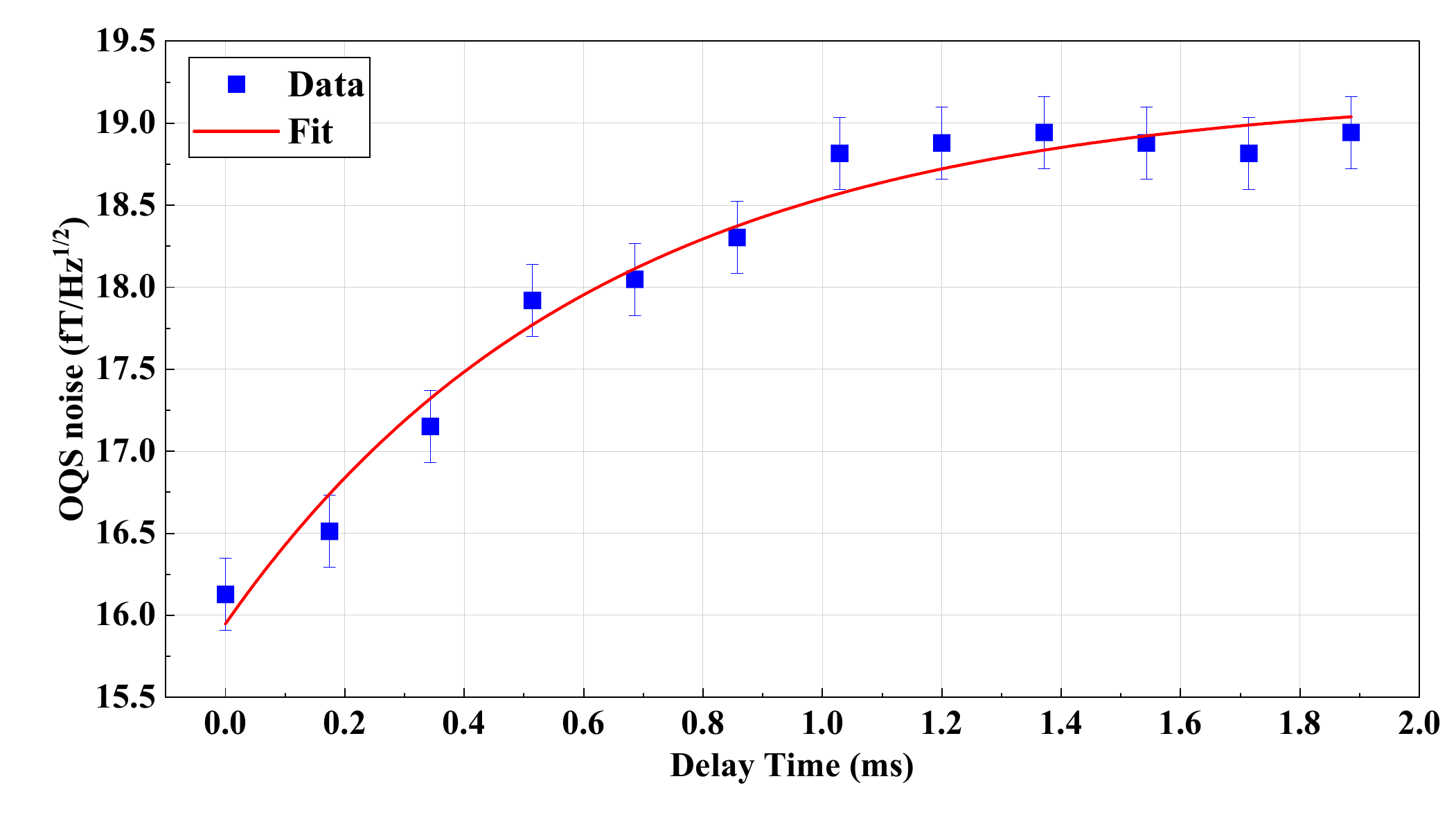}
\caption{\textbf{Correlated spin-noise measurements.}
Measured magnetic-field noise as a function of the delay time $\tau$ between the
preparation and readout probe pulse sequences. The symbols represent the
experimental data obtained using 40 averages. The solid curve is a fit to the
correlation model
$
N(\tau)=N_{\infty}-A\exp\!\left(-\frac{\tau}{T_2}\right),
$
where $N(\tau)$ is the measured noise, $N_{\infty} = (19.20\pm0.17)$ is the asymptotic noise, $A =(-3.25\pm0.18)$ is the correlated-noise reduction amplitude, and $T_2 = (0.62+0.10)$ is the transverse spin-relaxation time. Error bars represent the estimated experimental scatter obtained from the residuals of the fit.}
\label{fig:40 avr data}
\end{figure}

Figure~\ref{fig:40 avr data} shows the measured magnetic-field noise as a
function of the delay between the preparation and readout probe pulse
sequences. The noise increases monotonically with delay and approaches an
asymptotic value as the temporal spin correlations decay. The experimental
data are well described by the exponential model
\[
N(\tau)=N_{\infty}-A\exp(-\tau/T_2),
\]
yielding the transverse spin-relaxation time $T_2$ together with the
asymptotic noise level and the correlated-noise reduction amplitude.

The pulsed probe introduces residual artifacts associated with probe
modulation harmonics and probe-induced spin excitation. Although balanced
detection suppresses most of these components, residual correlated
contributions remain. These are substantially reduced by subtracting
consecutive waveforms acquired with synchronized phase cycling. As a result,
measurements performed with short acquisition windows ($\sim1$~ms), comparable
to the spin coherence time, exhibit a lower measured noise than measurements
performed at longer delay times.

\begin{figure} 
\centering \includegraphics[width=5 in]{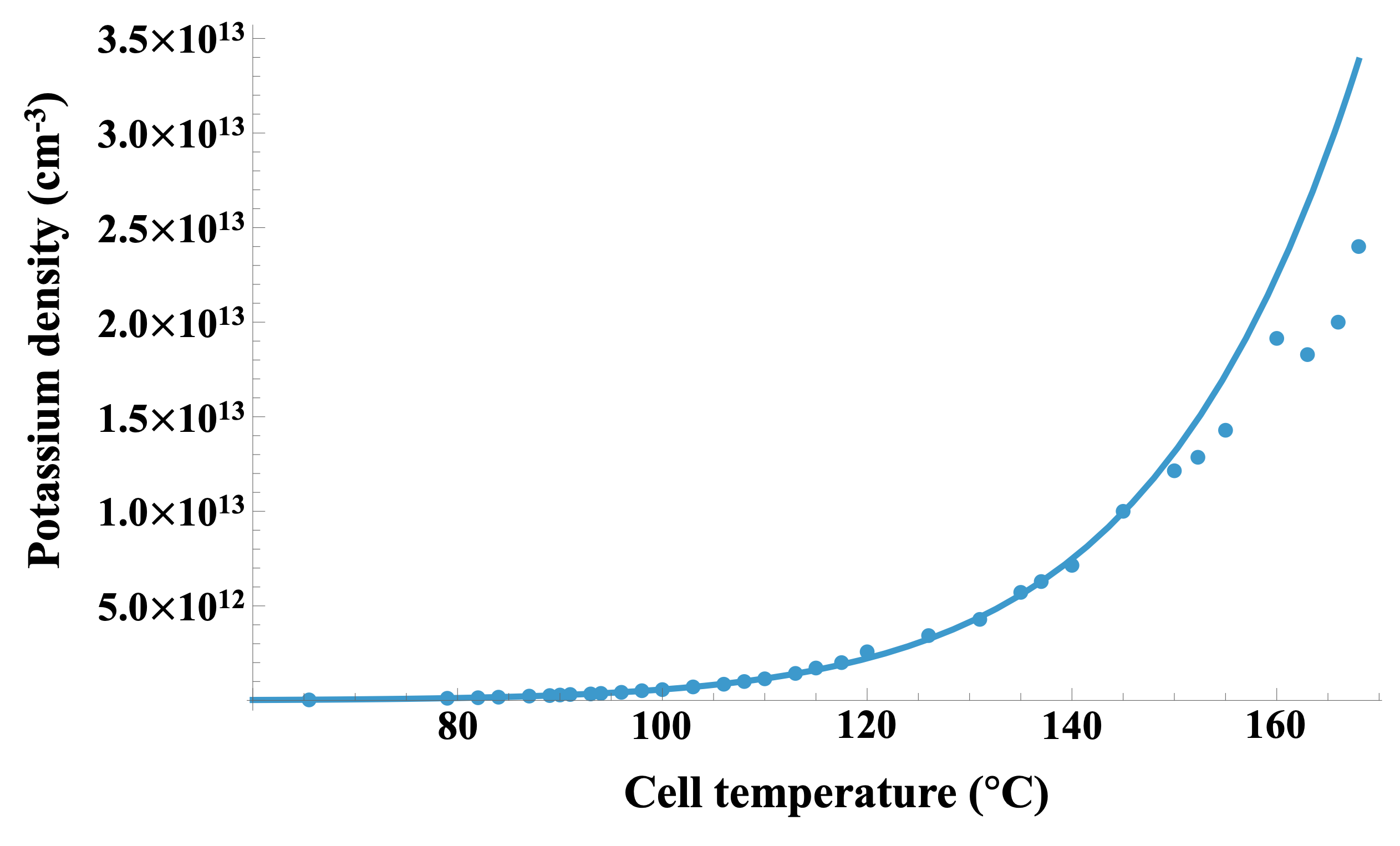} 
\caption{\textbf{Atomic density measurements using the adiabatic magnetometer signal.} The adiabatic magnetometer signal is proportional to the atomic density in the potassium vapor for a fixed probe laser detuning, which was monitored with a wavemeter. The coefficient of proportionality were adjusted to minimize the error between the measured signal and predicted saturation density curve (solid line) at the temperature below 110$^\circ$C. However, we find that above 150$^\circ$C the measured density significantly deviates from the saturation density prediction: at 168$^\circ$C, instead of $3.5\times 10^{13}$ it is only $2.5\times 10^{13}$. }
 \label{fig:dens fig} 
 \end{figure}
 
The atomic density in the potassium vapor cell was independently estimated
from the adiabatic magnetometer signal~\cite{SAVUKOVAQS}, as shown in
Fig.~\ref{fig:dens fig}. Using the measured potassium density, effective
interaction volume, and the measured transverse spin-relaxation time $T_2$,
the fundamental spin noise is estimated to lie in the range
$6.4$--$8.6~\mathrm{fT}/\sqrt{\mathrm{Hz}}$.
The probe-background noise was independently measured to be
$17~\mathrm{fT}/\sqrt{\mathrm{Hz}}$
with the pump beam blocked while retaining the same pulsed-probe sequence,
phase cycling, acquisition windows, subtraction procedure, and calibration
used for the data shown in Fig.~\ref{fig:40 avr data}.Assuming statistically independent spin and probe-background noise, the
expected long-delay noise is
\[
N_{\rm tot}=\sqrt{N_{\rm spin}^2+N_{\rm pr}^2}
=18.16\text{--}19.05~
\mathrm{fT}/\sqrt{\mathrm{Hz}},
\]
which agrees well with the fitted asymptotic value
$N_{\infty}=19.20\pm0.17~
\mathrm{fT}/\sqrt{\mathrm{Hz}}$.
At shorter delays, the measured noise is reduced below this asymptotic value,
consistent with temporal correlations preserved by the pulsed measurement
protocol. The exponential recovery toward the asymptotic noise level is
characterized by the transverse spin-relaxation time $T_2$, reflecting the
decay of temporal spin correlations between the preparation and readout probe
sequences.

The observed delay dependence demonstrates temporal correlations in the
measured spin fluctuations over time scales comparable to the transverse
spin-relaxation time. While such behavior is qualitatively consistent with
correlated spin dynamics, the present measurements do not uniquely distinguish
this mechanism from other correlated-noise processes associated with pulsed
probing, including residual probe-induced correlations, diffusion, and
technical noise. The results therefore demonstrate the effectiveness of the
implemented pulsed probing and subtraction protocol for reducing correlated
measurement noise while suppressing coherent technical noise.
\section*{Discussion}
The present experiments demonstrate that the combination of pulsed probing, phase cycling, and window-shifted subtraction substantially suppresses coherent probe-induced artifacts and reduces the measured spin-noise level. This reduction is observed for probe delays shorter than the characteristic spin relaxation time $T_2$, indicating temporal correlations in the measured signal that align with correlated spin fluctuations. These observations are consistent with correlated spin dynamics expected in spin-squeezing protocols; however, the present measurements do not uniquely distinguish this interpretation from other correlated-noise mechanisms, including probe-induced coherent spin dynamics and filtering associated with the subtraction protocol. Further quantitative analysis and independent determination of a spin-squeezing parameter will be required to establish the degree of spin squeezing. The present work therefore establishes a practical measurement protocol for future quantitative studies of spin squeezing, while independent determination of a spin-squeezing parameter remains the subject of future investigation.

Two complementary techniques were employed to suppress coherent technical noise: phase cycling and two-window subtraction with variable delay. Phase cycling, widely used in NMR, alternates the phase of sequential measurements and subtracts correlated waveforms, reducing coherent probe-induced spin excitations and light-shift artifacts by approximately one order of magnitude. In the pulsed-probe configuration, repeated probe pulses generate reproducible coherent responses that remain correlated between successive measurements. Phase cycling efficiently suppresses these correlated contributions while preserving the stochastic spin fluctuations. The two-window subtraction method further exploits temporal correlations by subtracting acquisition windows separated by a controlled delay. The delay is selected so that coherent sinusoidal components, such as ringing produced by periodic probe perturbations, have nearly the same phase in both windows and therefore cancel upon subtraction. Experimentally, this method provides an additional reduction of coherent signals by approximately one order of magnitude. The combination of phase cycling and two-window subtraction substantially suppresses coherent technical noise, enabling measurements closer to the intrinsic spin-noise limit.

The purpose of employing both techniques is to suppress coherent technical noise—primarily arising from probe-induced excitation of the spins—and thereby approach the sensitivity limit set by fundamental spin noise. The demonstrated protocol is directly applicable to RF magnetometry and pulsed NMR/NQR measurements, where phase cycling is routinely employed, and may also benefit emerging dark-matter detection schemes based on high-$Q$ resonant circuits, in which correlated spin noise can contribute to the detector noise floor. Suppression of coherent probe-induced artifacts therefore represents an important step toward operating such sensors closer to their intrinsic sensitivity limits.

While phase-cycling subtraction suppresses signals that remain phase-coherent between consecutive measurements, it can be compatible with detection of finite-coherence signals if the subtraction interval exceeds the signal coherence time. In this regime, signal contributions from successive measurements add incoherently rather than canceling, whereas reproducible technical artifacts remain correlated and are efficiently suppressed. The temporal control of noise demonstrated in this work—based on synchronized probing and selective measurement of temporally correlated spin fluctuations—therefore provides a pathway toward reducing sensor noise while maintaining sensitivity to externally generated signals with limited coherence.

The present experiment was intentionally configured with a narrow probe beam
to enhance the relative contribution of spin fluctuations within a small
active volume, rather than to maximize the absolute magnetic-field
sensitivity. For fixed atomic density, optical path length, probe detuning,
and probe intensity, increasing the probe-beam cross-sectional area increases
the optical-rotation signal in proportion to the number of interrogated atoms,
whereas the photon shot noise increases only as the square root of the
detected optical power. Consequently, the equivalent magnetic-field noise
associated with photon shot noise scales approximately as the inverse square
root of the probe area. The spin-noise and probe-back-action contributions
have the same overall $1/\sqrt{nV}$ scaling in the sensitivity model.
Therefore, with the optical path length held fixed, the total fundamental
noise is expected to scale approximately as $1/\sqrt{V}$.

Expanding the probe beam from the present area of approximately
$0.002~\mathrm{cm}^2$ to uniformly interrogate a
$2\times2~\mathrm{cm}^2$ cross section would therefore give the illustrative
scaling
\[
18~\mathrm{fT}/\sqrt{\mathrm{Hz}}
\times
\sqrt{\frac{0.002}{4}}
\approx
0.4~\mathrm{fT}/\sqrt{\mathrm{Hz}}.
\]
This projection assumes that the probe intensity, detuning, polarization,
optical path length, and relaxation conditions can be maintained over the
larger beam area. Changes in absorption, optical depth, or probe-induced
relaxation would require re-optimization of the probe conditions.

Compared with previous investigations of spin squeezing in cold-atom ensembles and optical cavities, which operate under carefully controlled QND conditions, the present work is performed in a high-density warm-vapor RF OQS under practical operating conditions where diffusion, spin exchange, and technical noise play important roles. Rather than demonstrating a quantitative degree of spin squeezing, the present work establishes a practical pulsed measurement protocol that combines synchronized probing, phase cycling, and time-correlated subtraction to suppress coherent probe-induced artifacts and reduce the measured spin-noise level. 
Thus, the broader significance of this work lies in demonstrating that coordinated phase cycling and temporally aligned subtraction can suppress coherent probe-induced noise by nearly two orders of magnitude, enabling operation closer to the fundamental spin-noise limit in a practically relevant RF OQS. The demonstrated measurement protocol provides a practical experimental framework for future investigations of quantum-enhanced sensing, including quantitative studies of spin squeezing in warm-vapor OQSs, and may also improve the performance of NMR, NQR, and related sensing applications.

\section*{Methods}

\subsection*{RF optical quantum sensor.}

The RF OQS was based on a potassium vapor cell with approximate dimensions of $2\times2\times2~\mathrm{cm}^3$. The cell was heated to $180^\circ$C using resistive heating with non-magnetic tungsten wire. During measurements, the heating current was periodically switched off to eliminate magnetic fields and magnetic noise associated with the heater current. The vapor cell was enclosed within a cylindrical ferrite shield to reduce ambient magnetic-field noise.
A circularly polarized pump beam propagated orthogonally to a linearly polarized probe beam. A static magnetic field applied along the pump-beam direction defined the RF resonance frequency and established the spin-precession axis. The probe beam diameter inside the cell was approximately $0.5~\mathrm{mm}$, measured using a pinhole method. The relatively small probe beam was intentionally chosen to enhance the relative contribution of intrinsic spin fluctuations within the active sensing volume. The probe laser frequency was detuned from the potassium D$_1$ transition to reduce probe-induced spin destruction while maintaining sufficient optical rotation signal.



\subsection*{Balanced optical detection.}
After passing through the vapor cell, the probe beam was analyzed using a polarizing beam splitter and detected with balanced photodetectors. The photocurrents were converted to voltages using a custom transimpedance amplifier and electronically subtracted to obtain the differential optical rotation signal.
Balanced detection was optimized by fine adjustment of the input polarizer angle to suppress common-mode intensity fluctuations and residual amplitude modulation introduced by the AOM. The differential signal was subsequently amplified prior to digitization.

\subsection*{Signal processing.}

The recorded waveforms were processed using two complementary
noise-suppression techniques: phase cycling and two-window subtraction.

In the phase-cycling stage, two consecutive waveforms were acquired with
opposite phases of the applied RF excitation. During processing, the sign of
the second waveform was selected according to the applied RF phase and the two
waveforms were averaged. For the desired RF signal, this phase correction
preserved the signal amplitude, while statistically independent noise was
reduced by a factor of $\sqrt{2}$. Conversely, coherent probe-induced
artifacts having the same phase in successive measurements were efficiently
suppressed. Experimentally, this behavior was verified by changing the sign of
the second recorded waveform in the processing routine and by reversing the
phase of a calibrated RF excitation.

The second processing stage employed two-window subtraction. Two acquisition
windows separated by a controllable delay were extracted from each
measurement, and their difference was calculated without additional
normalization. The delay was chosen so that coherent oscillatory components
produced by probe-induced perturbations possessed nearly identical phase in
the two windows and therefore canceled upon subtraction. In contrast, signal
components having opposite phase in the two windows added constructively,
while statistically independent noise increased only by $\sqrt{2}$, resulting
in an overall improvement of the signal-to-noise ratio by approximately
$\sqrt{2}$ relative to a single acquisition window.

The combination of phase cycling and window-shifted subtraction provided
nearly two orders of magnitude suppression of coherent technical noise,
including residual magnetic-field interference penetrating the ferrite shield,
while preserving sensitivity to appropriately phase-cycled external RF
signals. These processing techniques enabled measurements close to the
intrinsic spin-noise level.

\subsection*{Noise calibration.}
The conversion from OQS output to equivalent magnetic-field units was obtained using a calibrated RF magnetic field generated by an excitation coil. The resulting OQS response established the conversion factor between the measured differential voltage and the equivalent magnetic field. The calibration was performed using the same optical, electronic, and acquisition settings as the spin-noise measurements, but prior to phase cycling and window-shifted subtraction.

Phase cycling and window-shifted subtraction were introduced to suppress coherent probe-induced spin excitations while preserving the statistical properties of the fundamental spin fluctuations. In both procedures, two measurements are combined and normalized by a factor of $1/\sqrt{2}$. This normalization preserves the RMS level of statistically independent spin fluctuations and is therefore identical to the conventional spin-noise analysis used in RF atomic magnetometry~\cite{SavukovRFMag}. Consequently, in the absence of temporal correlations, the processed measurements yield the same spin-noise spectral density as the conventional analysis.
The two processing steps play different roles. Phase cycling combines consecutive measurements acquired with opposite probe phases. Since coherent probe-induced responses are reproducible between successive acquisitions, they cancel upon subtraction, while statistically independent spin fluctuations are preserved. In contrast, window-shifted subtraction combines two acquisition windows within the same measurement. When the delay between the windows is much longer than the transverse relaxation time $T_2$, the spin fluctuations become uncorrelated and the RMS noise is again preserved by the $1/\sqrt{2}$ normalization. For delays shorter than $T_2$, however, the spin fluctuations remain partially correlated, resulting in a reduction of the measured variance. Thus, any reduction of the measured noise originates from temporal correlations of the spin fluctuations rather than from the subtraction procedure itself.

For coherent externally applied RF signals, the subtraction procedures modify the signal transfer function according to the phase relationship between successive acquisitions. This behavior was characterized independently using calibrated RF excitations. Analogous to conventional phase cycling in NMR, synchronously reversing the phase of the applied RF excitation allows the desired signal to add constructively while coherent probe-induced artifacts cancel. Consequently, the subtraction protocol suppresses coherent technical noise without reducing the sensitivity to appropriately phase-cycled external RF signals.

Probe photon shot noise was estimated from the detected photocurrent with the pump laser beam off, $\sigma_I=\sqrt{2eI}$, and converted into voltage using the measured transimpedance gain. The estimate was verified experimentally by operating the vapor cell at low temperature, where the contribution of atomic spin noise becomes negligible.

The intrinsic spin-noise level was estimated independently from the measured atomic density, effective interaction volume, and transverse relaxation time using the conventional RF OQS analysis~\cite{SavukovRFMag}. These estimates were compared with the measured noise spectra to evaluate the degree of spin-noise reduction achieved using the pulsed measurement protocol.

The reported magnetic noise spectral densities were obtained from the fast Fourier transform (FFT) amplitudes after normalization by the square root of the FFT resolution bandwidth, which is determined by the acquisition time of the analyzed time-domain window. This FFT bandwidth should be distinguished from the intrinsic sensor bandwidth, which is determined by the transverse spin relaxation time. Under the operating conditions used in this work, the effective transverse relaxation time was approximately $T_2 \approx 1~\mathrm{ms}$, corresponding to a Lorentzian full width at half maximum of
\[
\Delta f_{\mathrm{FWHM}}=\frac{1}{\pi T_2}\approx320~\mathrm{Hz}.
\]
Additional broadening due to probe-induced spin destruction and residual magnetic-field gradients was relatively small, resulting in an overall bandwidth comparable to that reported previously for this RF OQS~\cite{SavukovRFMag}. Thus, the FFT resolution bandwidth determines the statistical uncertainty of the measured spectral density, whereas the intrinsic sensor bandwidth determines the frequency response of the sensor.
\bibliography{referencesn}

\section*{Author contributions}
I.S. and Y.J.K. equally performed the experiments, analyzed the data, and contributed to the manuscript.

\section*{Funding}
We acknowledge this work was supported by Quantum Information Science Enabled Discovery
2.0 (QuantISED 2.0) for the United States Department of Energy, Office of Science, Office of High Energy Physics (DE-FOA-0003354).

\section*{Declarations}

\section*{Data availability}
The datasets generated and analyzed during the current study are available from the corresponding author upon reasonable request.

\section*{Additional Information}
\textbf{Competing financial interests:} The authors declare no competing financial interests. 

\end{document}